\documentclass[12pt]{article}
\usepackage{amsfonts}
\usepackage{epsfig}
\usepackage{amsmath,verbatim,color,amssymb,epsfig,bm,mathrsfs,amsthm,latexsym,subfigure,graphicx}
\usepackage[round]{natbib}
\usepackage{hyperref}
\hypersetup{colorlinks=true,linkcolor=blue,citecolor=blue,urlcolor=blue}
\usepackage{array}
\usepackage{graphics}
\usepackage{subfigure}
\usepackage{caption}
\usepackage{multirow}

\usepackage{appendix}

\usepackage[figuresright]{rotating}

\usepackage{url}

\usepackage{amsmath,verbatim,color,amssymb,epsfig}
\usepackage{graphicx}
\usepackage{threeparttable}
\allowdisplaybreaks
\begin{document}

\makeatletter 
\@addtoreset{equation}{section}
\makeatother  

\def\yincomment#1{\vskip 2mm\boxit{\vskip 2mm{\color{red}\bf#1} {\color{blue}\bf --Yin\vskip 2mm}}\vskip 2mm}
\def\squarebox#1{\hbox to #1{\hfill\vbox to #1{\vfill}}}
\def\boxit#1{\vbox{\hrule\hbox{\vrule\kern6pt
          \vbox{\kern6pt#1\kern6pt}\kern6pt\vrule}\hrule}}

\def\theequation{\thesection.\arabic{equation}}
\newcommand{\ds}{\displaystyle}

\renewcommand{\abstractname}{Abstract}

\newcommand{\bcU}{\boldsymbol{\cal U}}
\newcommand{\bbeta}{\boldsymbol{\beta}}
\newcommand{\bdelta}{\boldsymbol{\delta}}
\newcommand{\bDelta}{\boldsymbol{\Delta}}
\newcommand{\boldeta}{\boldsymbol{\eta}}
\newcommand{\bxi}{\boldsymbol{\xi}}
\newcommand{\bfeta}{\boldsymbol{\eta}}
\newcommand{\bGamma}{\boldsymbol{\Gamma}}
\newcommand{\bSigma}{\boldsymbol{\Sigma}}
\newcommand{\balpha}{\boldsymbol{\alpha}}
\newcommand{\bOmega}{\boldsymbol{\Omega}}
\newcommand{\btheta}{\boldsymbol{\theta}}
\newcommand{\bmu}{\boldsymbol{\mu}}
\newcommand{\bnu}{\boldsymbol{\nu}}
\newcommand{\bgamma}{\boldsymbol{\gamma}}
\newcommand{\bpsi}{\boldsymbol{\psi}}
\newcommand{\bphi}{\boldsymbol{\phi}}
\newcommand{\bomega}{\boldsymbol{\omega}}
\newcommand{\comm}[1]{}

\newtheorem{defi}{\sc Definition}[section]
\newtheorem{theo}{\bf Theorem}
\newtheorem{lemm}{Lemma}
\newtheorem{rem}{Remark}
\newtheorem{coll}{\bf Corollary}


\setcounter{secnumdepth}{3}

\baselineskip=24pt
\begin{center}
{\Large \bf Bayesian Mendelian randomization with study heterogeneity and data partitioning for large studies}
\end{center}


\begin{center}
{\bf Linyi Zou, Hui Guo,{\footnote
{
Corresponding Author: Hui Guo (E-mail: hui.guo@manchester.ac.uk),
Centre for Biostatistics, The University of Manchester,
Manchester, UK.
}}
 and Carlo Berzuini}
\end{center}

\begin{center}
{\emph{Centre for Biostatistics, The University of Manchester, Jean McFarlane Building,
Oxford Road, Manchester M13 9PL, UK}}\\

\end{center}

\vspace{3mm}

\begin{center}
\begin{abstract}

\textbf{Background}: Mendelian randomization (MR) is a useful approach to causal inference from observational studies when randomised controlled trials are not feasible. 
However, study heterogeneity of two association studies required in MR is often overlooked. When dealing with large studies, recently developed Bayesian MR is limited by its computational expensiveness.

\textbf{Methods}: We addressed study heterogeneity by proposing a random effect Bayesian MR model with multiple exposures and outcomes. For large studies, we adopted a subset posterior aggregation method to tackle the problem of computation. In particular, we divided data into subsets and combine estimated subset causal effects obtained from the subsets''. The performance of our method was evaluated by a number of simulations, in which part of exposure data was missing.

\textbf{Results}: Random effect Bayesian MR outperformed conventional inverse-variance weighted estimation, whether the true causal effects are zero or non-zero. Data partitioning of large studies had little impact on variations of the estimated causal effects, whereas it notably affected unbiasedness of the estimates with weak instruments and high missing rate of data. Our simulation results indicate that data partitioning is a good way of improving computational efficiency, for little cost of decrease in unbiasedness of the estimates, as long as the sample size of subsets is reasonably large.

\textbf{Conclusions}:  We have further advanced Bayesian MR by including random effects to explicitly account for study heterogeneity. We also adopted a subset posterior aggregation method to address the issue of computational expensiveness of MCMC, which is important especially when dealing with large studies.
Our proposed work is likely to pave the way for more general model settings, as Bayesian approach itself renders great flexibility in model constructions.

\end{abstract}
\end{center}

\noindent
Abbreviations:



\vspace{0.5cm}




\section{Background}\label{sec1}
\indent
Mendelian randomization (MR) (\cite{Martjin1986, George2003, Debbie2008}) is a useful approach to causal inference from observational studies when randomised controlled trials are not feasible. 
It uses genetic variants as instrumental variables (IVs) to explore putative causal relationship between an exposure and an outcome.  Conventional MR methods (\cite{Toby2013, Jack2015, Jack2016, Zhao2018, Carlo2018, Stephen2014, Frank2003, Elinor2012}) have mainly used summary statistics of IV-exposure association and IV-outcome association analyses, from a single study (one-sample) or two independent studies (two-sample). Among recent developments of MR methods, a Bayesian approach  (\cite{Carlo2018}, \cite{Zou2020}) has been proposed to tackle overlapping samples in which a subset of participants are common in two association studies. This comes from the idea that overlapping-  and two- sample settings can be treated as problems of missing data, which can then be imputed through Markov chain Monte Carlo (MCMC) while estimating causal effects of interest. This way, we take full advantage of all the observed and imputed data. Bayesian MR also offers great flexibility of modelling complex data structure and explicitly quantifies uncertainties of model parameters.

It is not uncommon that studies from different research groups are designed to address similar (but not exactly the same) scientific questions. For example, in a genome-wide association study ($Study~1$), data of genetic variants and hypertension status (outcome) are collected to identify outcome-associated genetic variants. In another independent study ($Study~2$), besides this aim, the investigator is also interested in causal effect of blood pressure medication (exposure) on hypertension. Therefore, exposure information is also recorded. To investigate the exposure-outcome causal relationship, a conventional option would be one-sample MR using data from $Study~2$ only, without data from $Study~1$. Another option would be a two-sample MR which will use genetic variants and the outcome data from $Study~1$, and genetic variants and the exposure data from $Study~2$. In other words, the outcome data of $Study~2$ will be discarded. Both of the options will involve removal of data which, in our view, is not necessary. We would rather combine observed data from the two studies, and impute exposure data for $Study~1$ in a Bayesian MR model. However, it is well possible that the two studies are not homogenous, which should be taken into consideration in the model.

Another important aspect of Bayesian MR analysis (in fact, all kinds of data analysis) is computation, as we are in the era of big data. MCMC requires a large number of iterations and a complete scan of data for each iteration (\cite{Xue2019}). Thus, it is often computationally challenging, and sometimes even prohibitive. There is a need to address this issue in many research areas. An intuitive solution would be dividing data into a number of subsets and enabling data analysis in parallel.

This paper aims to address study heterogeneity and data partitioning for large studies in Bayesian MR. In Section \ref{sec2}, we build a Bayesian MR model including multiple IVs, exposures and outcomes based on two independent studies, of which one has exposure data completely missing. A random effect model is proposed to account for study heterogeneity. We adopt a data partitioning and subset posterior aggregation method (\cite{Xue2019}) for analysis of large studies. Simulation experiments are carried out for different configurations of IV strength and missing rate of exposure data. Section \ref{sec7} evaluates the performance of our proposed method, followed by discussion and conclusions in Section \ref{sec8}.

\section{Methods}\label{sec2}
\subsection{Bayesian MR with study heterogeneity}\label{sec2.1}
\indent

\begin{figure}[h!]
\centering
  \includegraphics[scale=0.5]{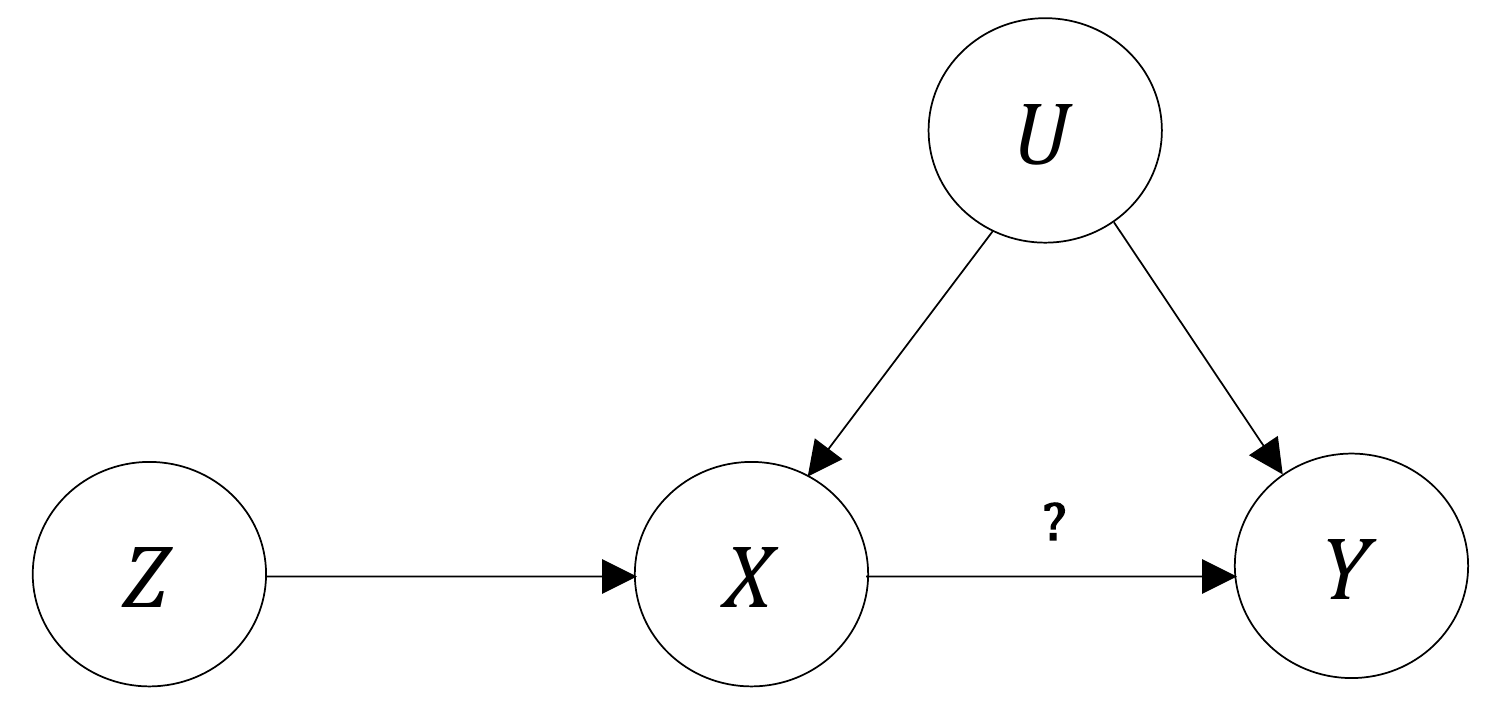}\\
  \caption{Schematic representation of the three assumptions required in Mendelian randomization.}
  \label{figure1}
\end{figure}

Let $X$ denote the exposure, $Y$ the outcome, and $U$ a set of unobserved confounders between $X$ and $Y$. Traditional MR (\cite{Stephen2014}) requires that an IV (denoted by $Z$) is : $i$) associated with the exposure $X$, $ii$) not associated with the confounders $U$, and $iii$) associated with the outcome $Y$ only through the exposure $X$. These three assumptions can be graphically expressed as Figure \ref{figure1} in which our interest is whether $X$ causes $Y$ (the $X \rightarrow Y$ arrow).

\comm{
\begin{itemize}
  \item[] $\rm A_{1}$: $Z$ is associated with $X$;
  \item[] $\rm A_{2}$: $Z$ is independent of $U$;
  \item[] $\rm A_{3}$: $Z$ is independent of $Y$, conditioning on $(X, U)$.
\end{itemize}

These assumptions are graphically expressed in Figure \ref{figure1}. The $Z \rightarrow X$ arrow denotes a non-zero association between the IV and the exposure, in accord with $\rm A_{1}$. Assumption $\rm A_{2}$ also follows from the graph. Also $\rm A_{3}$ follows from the graph (it would not if an arrow pointed directly from $Z$ to $Y$). The $X \rightarrow Y$ arrow represents the putative causal effect of $X$ on $Y$, that is, the object of our inference.
}

\comm{
\subsection{}\label{sec2.1}
\indent
}

\begin{figure}[h!]
\centering
  \includegraphics[scale=0.65]{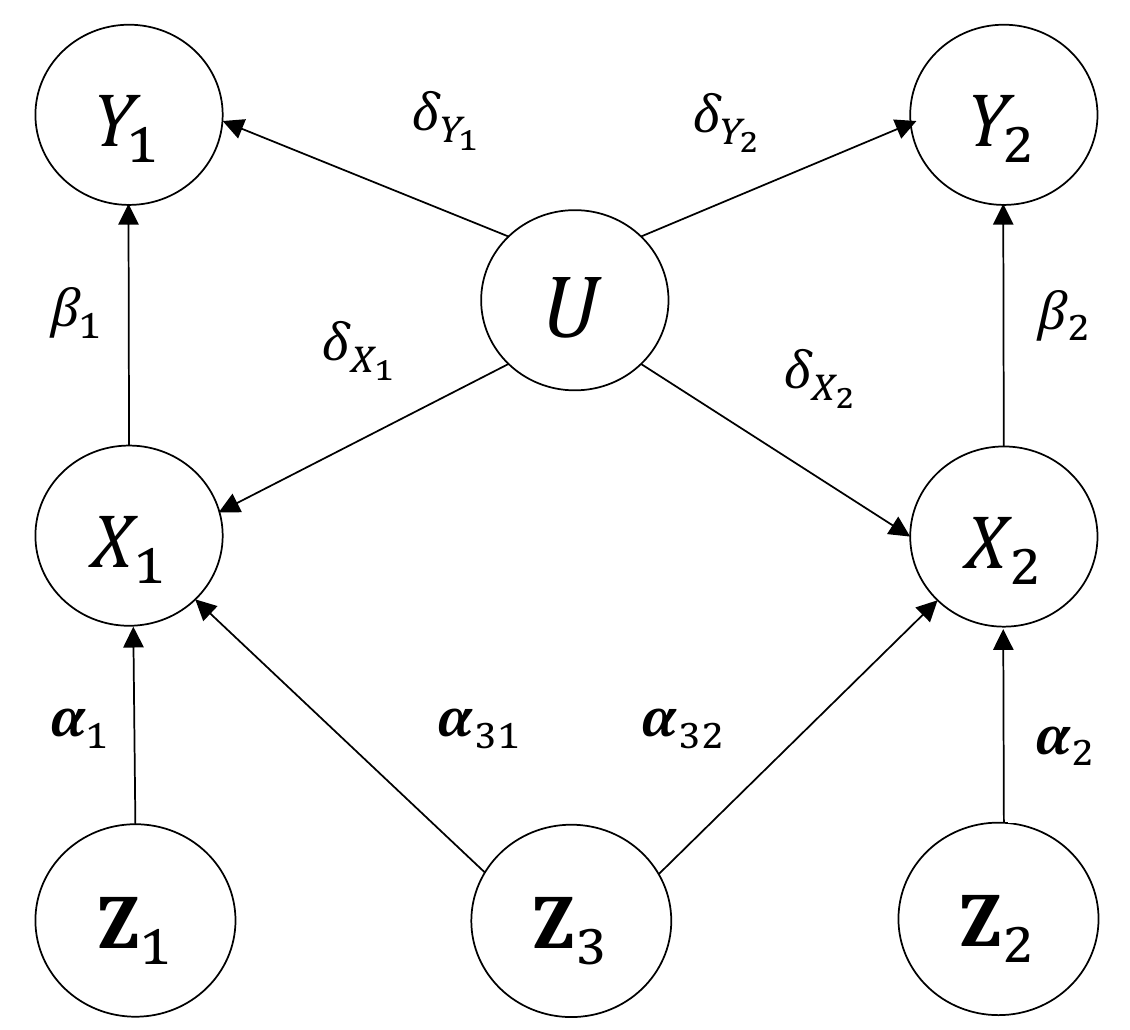}\\
  \caption{Graphical model of Mendelian randomisation with outcomes $Y_{1}$ and $Y_{2}$, exposures $X_{1}$ and $X_{2}$ and unobserved confounder $U$. $\mathbf{Z}_{1}$ consists of $L$ instrumental variables of $X_{1}$ and $\mathbf{Z}_{2}$ consists of $K$ instrumental variables of $X_{2}$. In addition, $\mathbf{Z}_{3}$ consists of $M$ instrumental variables shared between $X_{1}$ and $X_{2}$. The instrumental variables are assumed to be mutually independent.}
  \label{figure21}
\end{figure}

Without loss of generality, we consider a complex data generating process, as shown in Figure \ref{figure21}, involving three sets of IVs ($\mathbf{Z}_{1}$, $\mathbf{Z}_{2}$, $\mathbf{Z}_{3}$), where $\mathbf{Z}$ consists of $L, K, M$ independent IVs respectively, two exposures ($X_{1}$, $X_{2}$), two outcomes ($Y_{1}$, $Y_{2}$).

It has been shown that overlapping-sample and two-sample settings can be treated as problems of missing data in Bayesian MR, such that data imputation can be carried out based on observed data from two association studies, by assuming data was missing at random. This has led to improved precision of the estimated causal effect \cite{Zou2020}. However, when data was collected from different studies, the heterogeneity of studies should not be neglected.

Suppose we have data collected from two independent studies:
\begin{itemize}\label{1.2}
  \item $Study~A$ - observed data for IVs, exposures and outcomes $\{\mathbf{Z}_{1},\mathbf{Z}_{2},\mathbf{Z}_{3},X_{1},Y_{1},X_{2},Y_{2}\}$.
  \item $Study~B$ - observed data for IVs and outcomes $\{\mathbf{Z}_{1},\mathbf{Z}_{2},\mathbf{Z}_{3},Y_{1},Y_{2}\}$ only.
\end{itemize}
$Study~A$ includes fully observed data for MR, whereas $Study~B$ has exposure data completely missing. We shall include random effect terms in our MR model to capture study heterogeneity. By assuming standardised observed variables and linear additivity, according to Figure \ref{figure21}, our models are constructed as follows.

\vspace{2mm}

For $Study~A$,
\begin{eqnarray}\label{1.6}
U &\sim& N(0, 1),\\
  X_{1}  | \mathbf{Z}_{1},\mathbf{Z}_{3},U &\sim&
  N(\balpha_{1}\mathbf{Z}_{1} + \balpha_{31}\mathbf{Z}_{3} + \delta_{X_{1}}U, \sigma_{X_{1A}}^{2}),\\
\label{1.7}
  X_{2}  | \mathbf{Z}_{2},\mathbf{Z}_{3},U &\sim&
  N(\balpha_{2}\mathbf{Z}_{2} + \balpha_{32}\mathbf{Z}_{3} + \delta_{X_{2}}U, \sigma_{X_{2A}}^{2}),\\
\label{1.8}
  Y_{1} | X_{1},U &\sim& N(\beta_{1} X_{1}+\delta_{Y_{1}}U, \sigma_{Y_{1A}}^{2}),\\
\label{1.9}
  Y_{2} | X_{2},U &\sim& N(\beta_{2} X_{2}+\delta_{Y_{2}}U, \sigma_{Y_{2A}}^{2}).
\end{eqnarray}

For $Study~B$,
\begin{eqnarray}\label{1.10}
U &\sim& N(0, 1),\\
  X_{1}  | \mathbf{Z}_{1},\mathbf{Z}_{3},U &\sim&
  N(V_{X_{1}} + \balpha_{1}\mathbf{Z}_{1} + \balpha_{31}\mathbf{Z}_{3} + \delta_{X_{1}}U, \sigma_{X_{1B}}^{2}),\\
\label{1.11}
  X_{2}  | \mathbf{Z}_{2},\mathbf{Z}_{3},U &\sim&
  N(V_{X_{2}} + \balpha_{2}\mathbf{Z}_{2} + \balpha_{32}\mathbf{Z}_{3} + \delta_{X_{2}}U, \sigma_{X_{2B}}^{2}),\\
\label{1.12}
  Y_{1} | X_{1},U &\sim& N(V_{Y_{1}} + \beta_{1} X_{1}+\delta_{Y_{1}}U, \sigma_{Y_{1B}}^{2}),\\
\label{1.13}
  Y_{2} | X_{2},U &\sim& N(V_{Y_{2}} + \beta_{2} X_{2}+\delta_{Y_{2}}U, \sigma_{Y_{2B}}^{2}).
\end{eqnarray}

In the above pre-specified models, $\balpha$s are instrument strength parameters, and  $\delta$s are effects of $U$ on $X$s or $Y$s. Causal effects of $X$s on $Y$s are denoted by $\beta$s. The study heterogeneity is accounted for by $V$s. Note that $X_{1}$ and $X_{2}$ do not have observed data in $Study~B$, but they are part of data generating process, and thus, should be included in the model. $U$ is a sufficient scalar summary of the unobserved confounders. We assume that $U \sim N(0,1)$.

\comm{The parameters of this distribution are unidentifiable from the likelihood. However, under the Bayesian framework, non-identifiability can be negotiated by using a plausible prior which could induce a proper posterior on parameters of interest (\citet{Carlo2018}). All the exposures and outcome in this model are continuous.
}

\comm{
The causal effect of $X_{1}$ on $Y_{1}$ is denoted by $\beta_{1}$ and $X_{2}$ on $Y_{2}$ by $\beta_{2}$, the effects of the confounder $U$ on $X_{1}$, $X_{2}$, $Y_{1}$ and $Y_{2}$ as $\delta_{X_{1}}$, $\delta_{X_{2}}$, $\delta_{Y_{1}}$ and $\delta_{Y_{2}}$ respectively. The association from $\mathbf{Z}_{(\cdot)}$ to $X_{(\cdot)}$ is quantified by parameter $\alpha_{(\cdot)}$. For simplicity, we assume that all IVs are mutually independent and satisfy all the above stated MR assumptions, which is often referred to by calling them ``valid instruments".
}

\comm{
By assuming standardized observed variables and linear additive dependencies,  and in accord with the graph of Figure \ref{figure21}, we build our models as follows:
\begin{eqnarray}\label{1.1}
  U &\sim& N(0, 1),\\
\label{1.2}
  X_{1}  | \mathbf{Z}_{L},\mathbf{Z}_{M},U &\sim&
  N(\balpha_{L}\mathbf{Z}_{L} + \balpha_{M_{1}}\mathbf{Z}_{M} + \delta_{X_{1}}U, \sigma_{X_{1}}^{2}),\\
\label{1.3}
  X_{2}  | \mathbf{Z}_{K},\mathbf{Z}_{M},U &\sim&
  N(\balpha_{K}\mathbf{Z}_{K} + \balpha_{M_{2}}\mathbf{Z}_{M} + \delta_{X_{2}}U, \sigma_{X_{2}}^{2}),\\
\label{1.4}
  Y_{1} | X_{1},U &\sim& N(\beta_{1} X_{1}+\delta_{Y_{1}}U, \sigma_{Y_{1}}^{2}),\\
\label{1.5}
  Y_{2} | X_{2},U &\sim& N(\beta_{2} X_{2}+\delta_{Y_{2}}U, \sigma_{Y_{2}}^{2}),
\end{eqnarray}
where $N(a,b)$ stands for a normal distribution with mean $a$ and variance $b$. The $\sigma$s are standard deviations of independent random noise terms. The parameters of primary inferential interest are $\beta_{1}$ and $\beta_{2}$, which represent the causal effect of two exposures on two outcomes respectively. The $\balpha_{(\cdot)}$ parameter quantify the strengths of pairwise associations between the IV and the exposure. They are sometimes referred to as ``IV strengths", and should be significantly different from zero not to incur in the well known ``weak instrument bias" \cite{Burgess2011}. Finally, $U$ denotes an unobserved sufficient scalar summary of the unobserved confounders, which we set to be drawn from a $N(0,1)$ distribution, the parameters of this distribution being not identifiable from the likelihood, but, under the Bayesian framework, non-identifiability can be negotiated through using a plausible prior which could induce a proper posterior on parameters of our interest (\citet{Carlo2018}). All the exposures and outcome in this model are continuous. These models could be built on the basis of the Bayesian approachs developed by \cite{Carlo2018} and \cite{Zou2020}.
}

\comm{
\subsection{Random effect model}\label{sec2.2}
\indent

As mentioned in \cite{Zou2020}, an advantage of the Bayesian MR method is to incorporate previous studies to enhance statistical power and the precision of estimation. Considered that there are two data settings from two different studies, i.e. $Study~A$ and $Study~B$,
\begin{itemize}\label{1.2}
  \item $Study~A$: completed observation of $\{\mathbf{Z}_{L},\mathbf{Z}_{K},\mathbf{Z}_{M},X_{1},Y_{1},X_{2},Y_{2}\}$;
  \item $Study~B$: only $\{\mathbf{Z}_{L},\mathbf{Z}_{K},\mathbf{Z}_{M},Y_{1},Y_{2}\}$.
\end{itemize}
In $Study~A$, all values were completely observed, but in $Study~B$, due to some practical reasons, as mentioned in Section \ref{sec1}, we failed to get the information of exposures, $X_{1}$ and $X_{2}$ in our case. Precisely, $Study~B$ led to a setting with its exposures missing. In addition, due to the heterogeneity between $Study~A$ and $Study~B$, the difference between two studies should be included in the MR model. Thus, we turn to random effect model to take the heterogeneity into consideration.

According to the description above, we have the models as follows, in $Study~A$,
\begin{eqnarray}\label{1.6}
  X_{1}  | \mathbf{Z}_{L},\mathbf{Z}_{M},U &\sim&
  N(\balpha_{L}\mathbf{Z}_{L} + \balpha_{M_{1}}\mathbf{Z}_{M} + \delta_{X_{1}}U, \sigma_{X_{1A}}^{2}),\\
\label{1.7}
  X_{2}  | \mathbf{Z}_{K},\mathbf{Z}_{M},U &\sim&
  N(\balpha_{K}\mathbf{Z}_{K} + \balpha_{M_{2}}\mathbf{Z}_{M} + \delta_{X_{2}}U, \sigma_{X_{2A}}^{2}),\\
\label{1.8}
  Y_{1} | X_{1},U &\sim& N(\beta_{1} X_{1}+\delta_{Y_{1}}U, \sigma_{Y_{1A}}^{2}),\\
\label{1.9}
  Y_{2} | X_{2},U &\sim& N(\beta_{2} X_{2}+\delta_{Y_{2}}U, \sigma_{Y_{2A}}^{2}),
\end{eqnarray}
and in $Study~B$,
\begin{eqnarray}\label{1.10}
  X_{1}  | \mathbf{Z}_{L},\mathbf{Z}_{M},U &\sim&
  N(V_{X_{1}} + \balpha_{L}\mathbf{Z}_{L} + \balpha_{M_{1}}\mathbf{Z}_{M} + \delta_{X_{1}}U, \sigma_{X_{1B}}^{2}),\\
\label{1.11}
  X_{2}  | \mathbf{Z}_{K},\mathbf{Z}_{M},U &\sim&
  N(V_{X_{2}} + \balpha_{K}\mathbf{Z}_{K} + \balpha_{M_{2}}\mathbf{Z}_{M} + \delta_{X_{2}}U, \sigma_{X_{2B}}^{2}),\\
\label{1.12}
  Y_{1} | X_{1},U &\sim& N(V_{Y_{1}} + \beta_{1} X_{1}+\delta_{Y_{1}}U, \sigma_{Y_{1B}}^{2}),\\
\label{1.13}
  Y_{2} | X_{2},U &\sim& N(V_{Y_{2}} + \beta_{2} X_{2}+\delta_{Y_{2}}U, \sigma_{Y_{2B}}^{2}).
\end{eqnarray}
The heterogeneity will be depicted through $V_{X_{1}}$, $V_{X_{2}}$, $V_{Y_{1}}$ and $V_{Y_{2}}$ and be generated from the uniform distribution $U(-0.5,0.5)$ in the subsequent simulation experiment. Noted that $X_{1}$ and $X_{2}$ are unobservable in $Study~B$, but we still list the Models (\ref{1.10})-(\ref{1.13}) here for the following Bayesian imputation.
}

\comm{
\section{Methods}\label{sec3}
\indent

In this section, we use the Bayesian MR method, proposed in \cite{Carlo2018} and \cite{Zou2020}, that can be applied in the sample settings in which some of the individuals have missing exposure. More precisely, these individuals only have $\mathbf{Z}_{L}$ ,$\mathbf{Z}_{K}$, $\mathbf{Z}_{M}$, $Y_{1}$ and $Y_{2}$ (but not $X_{1}$ and $X_{2}$) measured. Therefore, it is more flexible than existing non-Bayesian methods in this respect, in which some information from data settings may be removed. In addition, this method can estimate $\beta_{1}$ and $\beta_{2}$ simultaneously, thus, take the correlations among different parameters into consideration.
}
\indent
The combined dataset of $Studies A$ and $B$ ($\mathcal{D}$, say) will contain fully observed data for the instruments and the outcomes. However, all participants in $Study~B$ have missing data of $X_{1}$ and $X_{2}$ which will be treated as unknown quantities and imputed from their conditional distributions given the observed data and current estimated parameters using MCMC. Let $X^{*}$ be imputed values of $X$. Our approach involves five steps as follows.
\begin{enumerate}

  \item Specify initial values for unknown parameters and the number of Markov iterations $T$.

  \item At the $t$th iteration, where $0 \leq t < T$, missing values of $X_{1}$ and $X_{2}$ in $Study~B$ will be filled with $X_{1}^{*}$ drawn from $N(V_{X_{1}}^{(t)} + \balpha_{1}^{(t)}\mathbf{Z}_{1} + \balpha_{31}^{(t)}\mathbf{Z}_{3} + \delta_{X_{1}}^{(t)}U, {\sigma_{X_{1B}}^{2}}^{(t)}$) and $X_{2}^{*}$ drawn from $N(V_{X_{2}}^{(t)} + \balpha_{2}^{(t)}\mathbf{Z}_{2} + \balpha_{32}^{(t)}\mathbf{Z}_{3} + \delta_{X_{2}}^{(t)}U, {\sigma_{X_{2B}}^{2}}^{(t)}$), respectively. $\mathbf{Z}_{1}$, $\mathbf{Z}_{2}$ and $\mathbf{Z}_{3}$ are observed values of IVs in $Study~B$. \comm{We use superscript ${(t)}$ for the random parameters at the $t$th iteration, with $t=0$ representing their initial values. }

  \item Create a single complete dataset including both the observed and the  imputed data.

  \item Estimate model parameters using MCMC and set $t \leftarrow t+1$.

  \item Repeat Steps 2-4 until $t=T$.

\end{enumerate}

Now we specify priors in the Bayesian model (\ref{1.6})-(\ref{1.13}). The priors of both $\beta_{1}$ and $\beta_{2}$ are set to a same distribution $N(0, 10^2)$, and those of IV strength parameters $\balpha$s are assumed to be independent and identically distributed: $\balpha_{1} \sim N_L(\mathbf{0}, 0.3^2\mathbf{I})$, $\balpha_{2} \sim N_K(\mathbf{0}, 0.3^2\mathbf{I})$, $\balpha_{31} \sim N_M(\mathbf{0}, 0.3^2\mathbf{I})$, and $\balpha_{32} \sim N_M(\mathbf{0}, 0.3^2\mathbf{I})$. Finally, we assign the priors of the standard deviations $\sigma$s to a same inverse-gamma distribution $ \emph{Inv-Gamma}(3, 2)$, and random effects $V$s to $N(0,1)$  in the Model (\ref{1.10})-(\ref{1.13}) for $Study~B$.

\subsection{Bayesian MR for large studies}\label{sec2.2}
\indent

Bayesian MR using MCMC is flexible in modelling complex data structure, quantifying uncertainties of parameters and enabling data imputation. However, it is computationally expensive and often requires a large amount of memory, especially for big data. It would be sensible to divide data $\mathcal{D}$ into a number of ($J$, say) subsets $D_1, D_2, ..., D_J$ with equal number of individuals. Bayesian MR can then be carried out in parallel based on these subsets, followed by aggregating posteriors obtained from each subset. Next, we will use a ``divide-and-combine" approach proposed by \cite{Xue2019} in our analysis.

For subset $D_{j}$, where $j=1, 2, ..., J$, let $\pi(\theta|D_{j})$ be posterior distribution of the parameters and $\mu^{(j)}$ mean vector of the posteriors. Let $\widehat{\mu} = \frac{1}{J}\sum_{j=1}^{J}\mu^{(j)}$ be the average of the mean vectors of the subset posteriors. According to \cite{Xue2019}, the posterior based on full data $\pi(\theta|\mathcal{D})$ can be estimated as the average of recentred subset posteriors.
\begin{eqnarray}\label{4.1}
  \widetilde{\pi}(\theta|\mathcal{D}) = \frac{1}{J}\sum\limits_{j=1}^{J}\widetilde{\pi}(\theta - \widehat{\mu} + \mu^{(j)}|D_{j}).
\end{eqnarray}
And it has been proved that (\cite{Xue2019})
\begin{equation}\label{4.2}
  E_{\widetilde{\pi}}(\theta) - E_{\pi}(\theta) = O_{p}(n_{j}^{-1}),
  \end{equation}
and
\begin{equation}\label{4.3}
  Var_{\widetilde{\pi}}(\theta) - Var_{\pi}(\theta) = o_{p}(n^{-1}),
\end{equation}
where $n_{j}$ is the sample size of the subsets and $n$ the sample size of the full dataset. $E_{\widetilde{\pi}}(\theta)$ and $E_{\pi}(\theta)$ are expectations of and $\pi(\theta|\mathcal{D})$ respectively. $Var_{\widetilde{\pi}}(\theta)$ and $Var_{\pi}(\theta)$ are their variances. It is easily seen that the difference in expectation depends on the sample size of the subsets and the difference in variation depends on the sample size of the full dataset.

\subsection{Simulations - Bayesian MR with study heterogeneity}\label{sec2.3}
\indent

We used simulated data to evaluate our Bayesian MR model with study heterogeneity in comparison with conventional MR methods. In particular, we considered 12 configurations including

\vspace{2mm}

\begin{itemize}

\item 3 missing rates of the exposures: 20\%, 50\%, 80\%

\item 2 degrees of the IV strength ($\balpha_1$,  $\balpha_2$, $\balpha_{31}$, $\balpha_{32}$): $\mathbf{0.1}$ and $\mathbf{0.3}$


\item Zero and non-zero causal effects of the exposures on the outcomes ($\beta_{1}, \beta_{2}$): 0 and 0.3.

\end{itemize}

The number of IVs was set to 15, 15 and 5 for $Z_1, Z_2$ and $Z_3$ respectively. Data of each IV were randomly drawn from a binomial distribution $B(2, 0.3)$ independently.  The specified values of the effects of $U$ on the exposures ($\delta_{X_{1}}, \delta_{X_{2}}$) and on the outcomes ($\delta_{Y_{1}}, \delta_{Y_{2}}$) were 1. Standard deviations $\sigma$s were set to 0.1. We simulated 200 datasets for each configuration.

For each dataset, we

\begin{itemize}
 \item simulated a dataset of sample size $n_A$ which contains observations of the IVs, exposures and outcomes (dataset $A$, denoted by $\mathcal{D}_A$);
 \item simulated a dataset of sample size $n_B$ which contains observations of the IVs, exposures and outcomes, then included data of the IVs and outcomes only as if the exposure data were missing (dataset $B$,  denoted by $\mathcal{D}_B$).
\end{itemize}

Sample size of $\mathcal{D}$, the combined data of $\mathcal{D}_A$ and $\mathcal{D}_B$, was set to 400 in all configurations, i.e., $n = n_A + n_B = 400$. The missing rate of the exposures was defined as  $\frac{n_B}{n}\times100\%$. For example, if missing rate was 50\%, we simulated $\mathcal{D}_A$ of sample size 200 and $\mathcal{D}_B$ of sample size 200. To allow for different degrees of study heterogeneity in different datasets, random effects $V$s in study $B$ were randomly drawn from a uniform distribution $U(0.5, 0.5)$ independently. Imputations of missing data and causal effect estimations were then performed simultaneously using MCMC in \texttt{Stan} (\cite{Stan2014, Martin2008}).

Estimated causal effects obtained from our Bayesian MR and two-sample inverse-variance weighted (IVW) estimation (\cite{Jack2016}) were compared using 4 metrics: mean, standard deviation (sd), coverage (proportion of the times that the 95\% credible/confidence intervals contained the true value of the causal effect) and power (proportion of the times that the 95\% credible/confidence intervals did not contain zero when the true causal effect was non-zero, only applicable when $\beta_{1} = \beta_{2} = 0.3$ by defination). Higher power indicates lower chance of getting false negative results. In IVW estimation, we used IV and outcome data from $\mathcal{D}_A$ and IV and exposure data from $\mathcal{D}_B$.

\subsection{Simulations - Bayesian MR with study heterogeneity for large studies}\label{sec2.4}
\indent

We also assessed the performance of dividing a big dataset into subsets in our Bayesian MR with study heterogeneity in simulation experiments. The simulation scheme was the same as above. However, the sample size of $\mathcal{D}$ was set to a much larger value 50,000. For each configuration, a single dataset was simulated by combining $\mathcal{D}_A$ and $\mathcal{D}_B$. We randomly divided data into 5 subsets of equal sample size, separately, for $\mathcal{D}_A$ ($\mathcal{D}_{A_1}, ..., \mathcal{D}_{A_5}$) and for $\mathcal{D}_B$ ($\mathcal{D}_{B_1}, ..., \mathcal{D}_{B_5}$).  Subset $\mathcal{D}_{i}$ was then constructed by combining $\mathcal{D}_{A_i}$ and $\mathcal{D}_{B_i}$, where $i=1, ..., 5$. This is to ensure that subset $\mathcal{D}_{i}$ had the same missing rate as that of the full data $\mathcal{D}$. Causal effects were estimated using $\mathcal{D}$, and using the 5 subsets in Bayesian MR. To explore the impact of different data partitioning strategies on estimated causal effects, we carried out the same analysis by also dividing data into 50 subsets of sample size 1,000.

\section{Results}\label{sec7}

\subsection{Simulation results - Bayesian MR with study heterogeneity}\label{sec3.1}
\indent

Table \ref{table1} displays simulation results when the true causal effects were non-zero ($\beta_{1}=\beta_{2}=0.3$). Each row of the table corresponds to a configuration of a specified missing rate and a degree of IV strength $\balpha$. Columns are estimated causal effects of $X_1$ on $Y_1$ ($\hat{\beta_{1}}$)  and of $X_2$ on $Y_2$ ($\hat{\beta_{2}}$) from our Bayesian method and from the IVW method evaluated using the four metrics. 
Unsurprisingly, the estimated causal effect of $X_1$ on $Y_1$  was very similar to that of $X_2$ on $Y_2$ in each configuration from Bayesian MR, because their true values were set to be the same and the model had a symmetrical structure, as shown in Figure \ref{figure21}. This was also observed in the results from the IVW method. However, Bayesian MR outperformed IVW uniformly across all the configurations, with less bias, higher precision, coverage and power. The impact of low missing rate was positive on coverage but negative on power in IVW. However, such impact was negligible in Bayesian MR. This was mainly due to much higher variations of the estimates, and consequently, much wider confidence intervals in IVW estimation. Weaker IVs had little influence on unbiasedness of the estimates and power, but resulted in slightly lower precision and coverage in Bayesian MR. However, there was a remarkable decrease in unbiasedness, precision and power as IV strength decreased.

Table \ref{table2} presents simulation results when the true causal effects were zero ($\beta_{1}=\beta_{2}=0$). Again, the results of $\hat{\beta_{1}}$ was very similar to those of $\hat{\beta_{2}}$ in each configuration, separately, from Bayesian MR and from IVW. Overall, both methods performed well.  However, Bayesian MR still outperformed IVW across all the configurations, with higher coverage and precision and less biased estimates. In both MR methods, missing rate did not have a notable effect on the estimates, whereas weaker IVs led to lower precision.

\begin{sidewaystable}[h!]
\scriptsize
  \centering
  \caption{Causal effects estimated from 200 simulated datasets for each configuration from two MR methods (Bayesian, IVW) when $\beta_{1}=\beta_{2}=0.3$, using four metrics: mean, standard deviation (sd), coverage and power. The six configurations were generated from three missing rates of the exposures (80\%, 50\%, 20\%) and two levels of IV strength ($\balpha = \mathbf{0.3}$ and $\mathbf{0.1}$). $\hat{\beta_{1}}$: estimated causal effect of $X_1$ on $Y_1$, $\hat{\beta_{2}}$:  estimated causal effect of $X_2$ on $Y_2$.}

    \begin{tabular}{|c|c|cccc|cccc|cccc|cccc|}
    \hline
    \multicolumn{1}{|c|}{\multirow{3}[3]{*}{\textbf{Missing rate}}} & \multirow{3}[3]{*}{\textit{$\balpha$}} & \multicolumn{8}{c|}{\textit{$\widehat{\beta_{1}}$}}                           & \multicolumn{8}{c|}{\textit{$\widehat{\beta_{2}}$}} \\
\cline{3-18}          &       & \multicolumn{4}{c|}{\textbf{Bayesian}} & \multicolumn{4}{c|}{\textbf{IVW}} & \multicolumn{4}{c|}{\textbf{Bayesian}} & \multicolumn{4}{c|}{\textbf{IVW}} \\
\cline{3-18}          &       & \textbf{mean} & \textbf{sd} & \textbf{coverage} & \textbf{power} & \textbf{mean} & \textbf{sd} & \textbf{coverage} & \textbf{power} & \textbf{mean} & \textbf{sd} & \textbf{coverage} & \textbf{power} & \textbf{mean} & \textbf{sd} & \textbf{coverage} & \textbf{power} \\
    \hline
    80\%  & 0.3   & 0.299  & 0.005  & 0.980  & 1     & 0.217  & 0.101  & 0.790  & 0.685 & 0.298  & 0.005  & 0.970  & 1     & 0.209  & 0.086  & 0.765  & 0.665 \\
          & 0.1   & 0.298  & 0.015  & 0.975  & 1     & 0.081  & 0.141  & 0.695  & 0.065 & 0.299  & 0.015  & 0.985  & 1     & 0.071  & 0.146  & 0.690  & 0.045 \\
    \hline
    50\%  & 0.3   & 0.300  & 0.004  & 0.975  & 1     & 0.245  & 0.118  & 0.920  & 0.580  & 0.299  & 0.004  & 0.980  & 1     & 0.265  & 0.113  & 0.935  & 0.595 \\
          & 0.1   & 0.302  & 0.013  & 0.960  & 1     & 0.169  & 0.277  & 0.925  & 0.115 & 0.302  & 0.013  & 0.955  & 1     & 0.122  & 0.268  & 0.900  & 0.075 \\
    \hline
    20\%  & 0.3   & 0.299  & 0.004  & 0.970  & 1     & 0.260  & 0.203  & 0.915  & 0.255 & 0.299  & 0.004  & 0.970  & 1     & 0.276  & 0.185  & 0.955  & 0.285 \\
          & 0.1   & 0.303  & 0.012  & 0.955  & 1     & 0.193  & 0.439  & 0.945  & 0.050  & 0.302  & 0.012  & 0.950  & 1     & 0.181  & 0.469  & 0.945  & 0.070 \\
    \hline
    \end{tabular}%
  \label{table1}%
\end{sidewaystable}

\begin{sidewaystable}[h!]
\scriptsize
  \centering
  \caption{Causal effects estimated from 200 simulated datasets for each configuration from two MR methods (Bayesian, IVW) when $\beta_{1}=\beta_{2}=0$, using four metrics: mean, standard deviation (sd), coverage and power. The six configurations were generated from three missing rates of the exposures (80\%, 50\%, 20\%) and two levels of IV strength ($\balpha = \mathbf{0.3}$ and $\mathbf{0.1}$). $\hat{\beta_{1}}$: estimated causal effect of $X_1$ on $Y_1$, $\hat{\beta_{2}}$:  estimated causal effect of $X_2$ on $Y_2$.}
    \begin{tabular}{|c|c|ccc|ccc|ccc|ccc|}
    \hline
    \multicolumn{1}{|c|}{\multirow{3}[3]{*}{\textbf{Missing rate}}} & \multirow{3}[3]{*}{\textit{$\balpha$}} & \multicolumn{6}{c|}{\textit{$\widehat{\beta_{1}}$}}           & \multicolumn{6}{c|}{\textit{$\widehat{\beta_{2}}$}} \\
\cline{3-14}          &       & \multicolumn{3}{c|}{\textbf{Bayesian}} & \multicolumn{3}{c|}{\textbf{IVW}} & \multicolumn{3}{c|}{\textbf{Bayesian}} & \multicolumn{3}{c|}{\textbf{IVW}} \\
\cline{3-14}          &       & \textbf{mean} & \textbf{sd} & \textbf{coverage} & \textbf{mean} & \textbf{sd} & \textbf{coverage} & \textbf{mean} & \textbf{sd} & \textbf{coverage} & \textbf{mean} & \textbf{sd} & \textbf{coverage} \\
    \hline
    80\%  & 0.3   & -0.001  & 0.005  & 0.960  & 0.007  & 0.061  & 0.955  & -0.001  & 0.005  & 0.955  & -0.005  & 0.062  & 0.960  \\
          & 0.1   & 0.004  & 0.016  & 0.960  & -0.010  & 0.112  & 0.965  & 0.004  & 0.015  & 0.960  & -0.001  & 0.130  & 0.960  \\
    \hline
    50\%  & 0.3   & 0.000  & 0.005  & 0.975  & -0.014  & 0.087  & 0.935  & 0.000  & 0.005  & 0.955  & -0.002  & 0.090  & 0.955  \\
          & 0.1   & 0.004  & 0.013  & 0.970  & 0.005  & 0.188  & 0.960  & 0.004  & 0.013  & 0.955  & -0.011  & 0.202  & 0.950  \\
    \hline
    20\%  & 0.3   & 0.000  & 0.004  & 0.950  & 0.010  & 0.148  & 0.930  & 0.000  & 0.004  & 0.965  & -0.003  & 0.152  & 0.935  \\
          & 0.1   & 0.003  & 0.012  & 0.965  & 0.012  & 0.394  & 0.920  & 0.003  & 0.012  & 0.965  & 0.020  & 0.361  & 0.945  \\
    \hline
    \end{tabular}%
  \label{table2}%
\end{sidewaystable}

\subsection{Simulation results - Bayesian MR with study heterogeneity for large studies}\label{sec3.2}
\indent

\begin{figure}[h!]
\centering
  \includegraphics[scale=0.3]{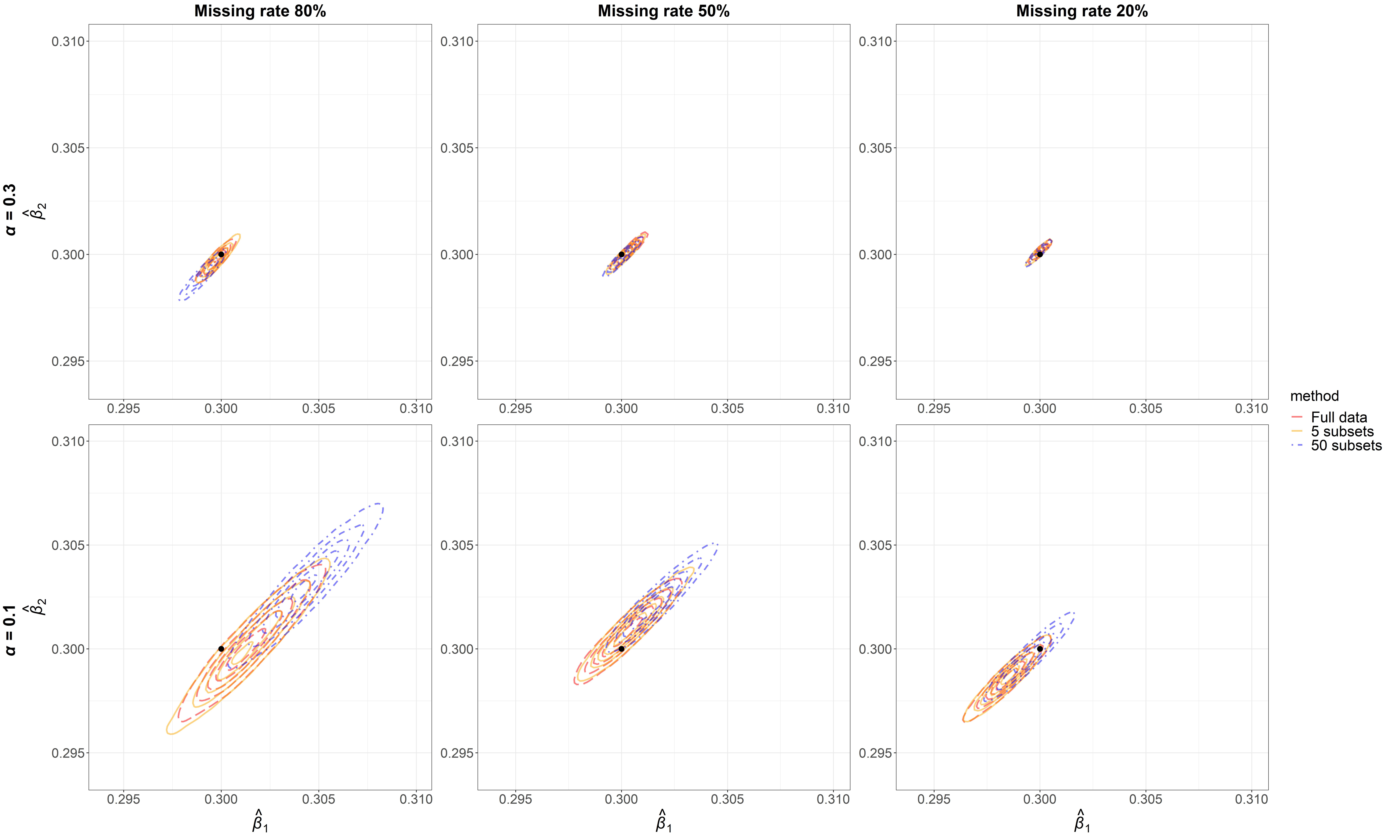}\\
  \caption{2D density contour based on simulated data with continuous $Y_{1}$ and $Y_{2}$ using our Bayesian method with subset posterior aggregation when $\beta_{1}=\beta_{2}=0.3$ and $\balpha = \mathbf{0.3}$ and $\mathbf{0.1}$.}
  \label{figure2}
\end{figure}

Figure \ref{figure2} depicts the joint posterior distributions of $\hat{\beta_{1}}$ (horizontal axis) and $\hat{\beta_{2}}$ (vertical axis) based on simulated data when the true causal effects were non-zero. Columns corresponds to three missing rates and rows two levels of IV strength. In each panel, the black dot denotes the values of true causal effects ($\beta_{1} = \beta_{2} = 0.3$). The red, orange and blue contours are 2-dimensional Gaussian kernel density estimation of the joint posterior (GKDEJP) from the full dataset, aggregated GKDEJP from five subsets and aggregated GKDEJP from fifty subsets respectively.  When IVs were strong in Bayesian MR analysis (top panels), estimated causal effects were close to their true values, with or without data partitioning. When IVs became weaker (bottom panels), the results from the full data were concordant with those from 5 subsets, but notably different from those based on 50 subsets. The impact of data partitioning was substantial with weak IVs and high missing rate.  This could be explained by Equation (\ref{4.2}), in which difference in mean of the GKDEJPs depends on the subset sample size $n_{j}$.
Difference in variance of the GKDEJPs was, however, not evident in the three sets of contours in each configuration, because it only depends on the sample size of the full data (Equation (\ref{4.3})) which was a fixed value 50,000. Our simulation results suggest that, in Bayesian MR with a large sample size, there is a trade-off between data partitioning for more efficient computations, and large enough sample size of each subset for preventing estimates from a decrease in unbiasedness.

\begin{figure}[h!]
\centering
  \includegraphics[scale=0.3]{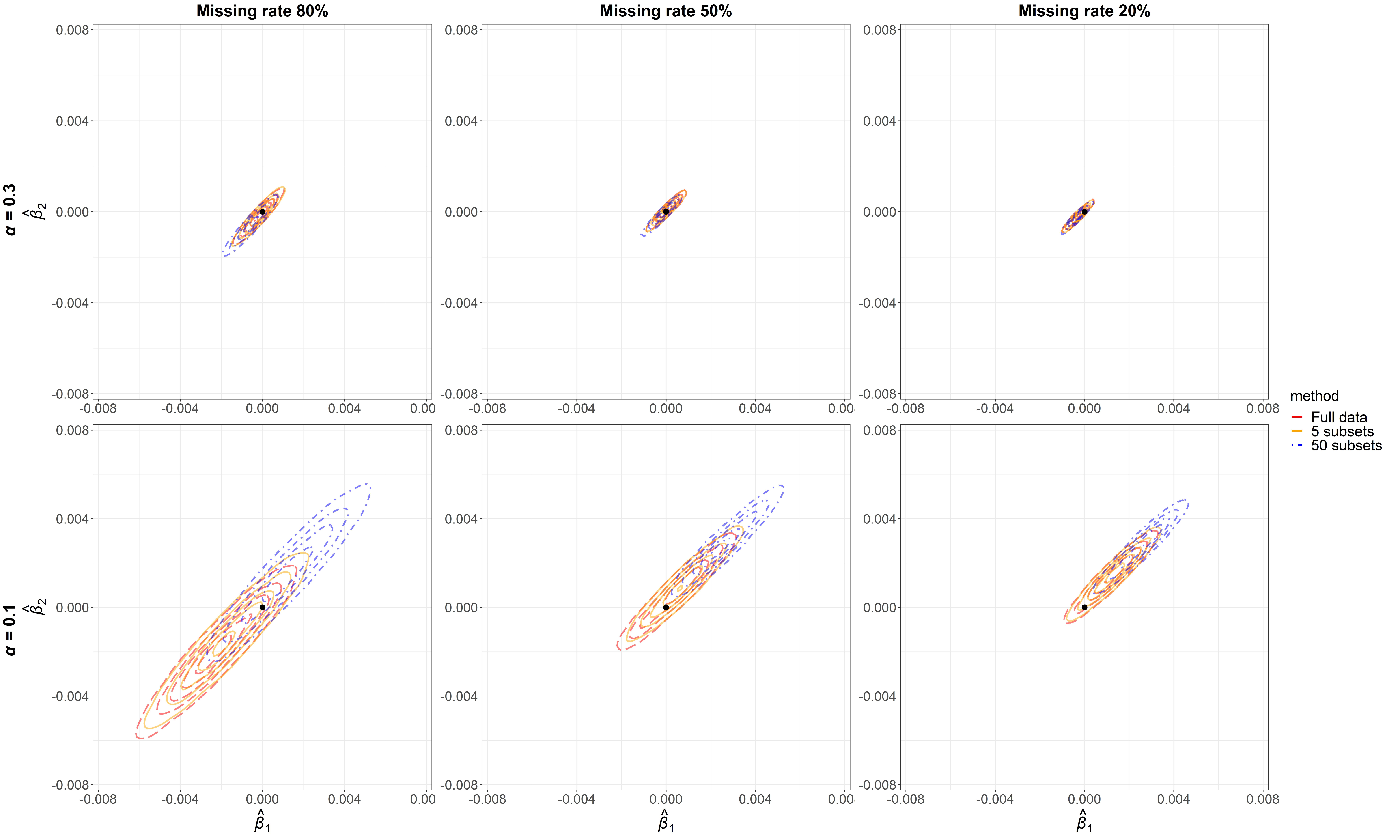}\\
  \caption{2D density contour based on simulated data with continuous $Y_{1}$ and $Y_{2}$ using our Bayesian method with subset posterior aggregation when $\beta_{1}=\beta_{2}=0$ and $\balpha = \mathbf{0.3}$ and $\mathbf{0.1}$.}
  \label{figure3}
\end{figure}

The same plots were presented in Figure \ref{figure3} when the true causal effects were zero. The performances of the three data partition strategies were very similar to those when the true causal effects were non-zero.

\section{Discussion and conclusions}\label{sec8}
\indent

Numerous MR methods have been developed in recent years. To the best of our knowledge, little attention has been focused on study heterogeneity. In this study, we have further advanced our Bayesian MR method (\cite{Carlo2018}, \cite{Zou2020}) by including random effects to explicitly account for study heterogeneity. \comm{This would be a   the random effect terms was integrated into the model to depict the heterogeneity between different studies. This is an important addition to the existing MR literature in which almost all of researchers only focused on a single study or simply neglected the heterogeneity among different studies.} We also adopted a subset posterior aggregation method proposed by \cite{Xue2019} to address the issue of computational expensiveness of MCMC, which is important especially when dealing with large studies. Our simulation results have indicated that the ``divide (data) and combine (estimated subset causal effects)'' is a good way of improving computational efficiency, for little cost of decrease in unbiasedness of the estimated causal effects, as long as the sample size of subsets is reasonably large.

A limitation of our method is that the analysis was carried out from simulated data based on a simple model with a small number of variables, given complex data generating process in real life. This study is also limited by analysis of a moderate number of configurations. Nevertheless, our proposed work is likely to pave the way for more general model settings, as Bayesian approach itself renders great flexibility in model constructions.


\clearpage
\phantomsection
\addcontentsline{toc}{section}{Bibliography}

\bibliography{bibliography}

\comm{

\clearpage
\phantomsection
\addcontentsline{toc}{section}{Appendix}

\begin{appendices}

\appendix

\begin{center}

\section*{Appendix}

\end{center}

\subsection*{1. Simulation experiment for our Bayesian method}
\indent

In our simulations, according to Models (\ref{1.6})-(\ref{1.13}), we consider a total of 12 configurations including
\vspace{2mm}

\begin{itemize}

\item 3 percentages of miss exposure (80\%, 50\%, 20\%)

\item 2 IV strengths: $\balpha = \mathbf{0.3}$ and $\mathbf{0.1}$


\item 2 levels of causal effects of $X_{1}$ on $Y_{1}$ and $X_{2}$ on $Y_{2}$: $\beta_{1} = \beta_{2} = 0.3$ and $0.$

\end{itemize}

For simplicity, the effect of $U$ on $X_{1}$, $X_{2}$, $Y_{1}$ and $Y_{2}$ is set to 1 ($\delta_{X_{1}} = \delta_{X_{2}} = \delta_{Y_{1}} = \delta_{Y_{2}} = 1$). We set $l = k = 15$ and $m = 5$. In each configuration, we simulated 200 datasets, step by step, as follows.

\vspace{2mm}

\begin{enumerate}
\item Generate a dataset $\textbf{H}$ which contains observed IVs ($\mathbf{Z}_{L}$ ,$\mathbf{Z}_{K}$ and $\mathbf{Z}_{M}$), exposures ($X_{1}$ and $X_{2}$) and outcome ($Y_{1}$ and $Y_{2}$) from 1000 independent individuals.
 \item Randomly sample $n_A$ individuals without replacement from $\textbf{H}$ and take their observations of all IVs, exposures and outcomes as dataset $A$;
 \item Randomly Sample $n_B$ individuals without replacement from $\textbf{H}-A = \{x \in \textbf{H},  x \notin A\}$ and take their observations of only IVs and outcomes as dataset $B$.

\end{enumerate}

\vspace{2mm}

\subsection*{2. Comparison of the performance of different methods}
\indent

The performance of our Bayesian method and other three MR methods was evaluated using 4 metrics:

\vspace{2mm}

\begin{itemize}

  \item mean (posterior mean)
  \item sd (standard deviation)
  \item coverage (the proportion of the times that the 95\% credible interval contained the true value of the causal effect)
  \item power (the proportion of the times that the 95\% credible interval did not contain value zero when the true causal effect was non-zero).

\end{itemize}

\vspace{2mm}

The causal effects of $X_{1}$ and $X_{2}$ on the outcome $Y_{1}$ and $Y_{2}$ (denoted as $\hat{\beta_{1}}$ and $\hat{\beta_{2}}$ respectively) were estimated from data simulated under the alternative hypothesis that $\beta_{1} = \beta_{2} = 0.3$, and from data simulated under the null hypothesis that $\beta_{1} = \beta_{2} = 0$. Note that the metric power was only applicable under the alternative hypothesis when $\beta_{(\cdot)} \neq 0$, by its definition. A high value of power indicates high sensitivity of the model to deviations of the data from the null or, equivalently, a low expected probability of false negatives.

We compared our method with the inverse-variance weighted (IVW) estimation (\cite{Jack2016}). \cite{Burgess2016} has shown that sample overlap increases type I error and leads to bias in classic MR methods. Thus, for IVW, we only included data of non-overlapping samples to make it a two-sample MR, precisely, the ``$\mathbf{Z}-\mathbf{X}$'' association from $A$ and the ``$\mathbf{Z}-\mathbf{Y}$'' association from $B$.

\subsection*{3. Simulation results}
\indent

Tables \ref{table1}-\ref{table2} display the summary of results obtained from the simulated data under the alternative hypothesis ($\beta_{1}=\beta_{2}=0.3$) and under the null hypothesis ($\beta_{1}=\beta_{2}=0$) respectively. Each row of the table corresponds to a configuration of specified percentage of missing exposures and IV strength $\balpha$. Columns are values of the four metrics of the estimated causal effects $\hat{\beta_{1}}$ and $\hat{\beta_{2}}$ obtained from our Bayesian method and classic IVW method.

Let us firstly focus on our Bayesian method in Table \ref{table1}. As $\balpha$ increased, the average of estimated causal effects (mean) became closer to the true values $\mathbf{0.3}$ generally and the variation (sd) decreased, concluding that higher IV strength reduced bias and resulted in more precise estimates. We note that, there is a trend of decrease in variation (sd) with the decrease of missing rate of exposures. This could be easily interpreted by the fact that a higher percentage of missing exposures tends to increase the uncertainty for the estimation. Compared with Bayesian method, the classic IVW method has a bad performance under the alternative hypothesis. As shown in Table \ref{table1}, the estimated means in the classic columns are largely deviated from $\mathbf{0.3}$, especially when the IV strength is weak ($=\mathbf{0.1}$). Classic method performs well under the null hypothesis when $\beta_{1}=\beta_{2}=0$ in terms of estimated mean (Table \ref{table2}), but across two tables, classic method led to a larger variation (sd) than our Bayesian method, meaning our method has a higher precision than the classic method.

The Bayesian approach results in a high statistical power (=1) and coverage under the alternative ($\geq$ 0.95) in all configurations (Table \ref{table1}), which outperforms classic method, especially in power. The coverage under the null in Table \ref{table2} from both two approaches are good ($\geq$ 0.92). In addition, it is interesting that the estimated results for $\beta_{1}$ and $\beta_{2}$ are very similar, this is because the model in this simulation experiment has a symmetrical structure, as shown in Figure \ref{figure21}.

\begin{sidewaystable}[h!]
\scriptsize
  \centering
  \caption{Causal effects estimated from simulated data with continuous $Y_{1}$ and $Y_{2}$ using two different methods (our Bayesian method and classic IVW method) when $\beta_{1}=\beta_{2}=0.3$. Mean, standard deviation (sd), coverage under the alternative (coverage) and power are displayed for the estimated causal effects of the exposures $X_{1}$ and $X_{2}$ on the outcomes $Y_{1}$ and $Y_{2}$. There are 6 configurations containing three percentages of missing $X$ (80\%, 50\%, 20\%) and two levels of IV strengths ($\balpha = \mathbf{0.3}$ and $\mathbf{0.1}$). The effects of $U$ on all the exposures and outcomes are set to 1.}
    \begin{tabular}{|c|c|cccc|cccc|cccc|cccc|}
    \hline
    \multicolumn{1}{|c|}{\multirow{3}[3]{*}{\textbf{Missing rate}}} & \multirow{3}[3]{*}{\textit{$\balpha$}} & \multicolumn{8}{c|}{\textit{$\widehat{\beta_{1}}$}}                           & \multicolumn{8}{c|}{\textit{$\widehat{\beta_{2}}$}} \\
\cline{3-18}          &       & \multicolumn{4}{c|}{\textbf{Bayesian}} & \multicolumn{4}{c|}{\textbf{Classic}} & \multicolumn{4}{c|}{\textbf{Bayesian}} & \multicolumn{4}{c|}{\textbf{Classic}} \\
\cline{3-18}          &       & \textbf{mean} & \textbf{sd} & \textbf{coverage} & \textbf{power} & \textbf{mean} & \textbf{sd} & \textbf{coverage} & \textbf{power} & \textbf{mean} & \textbf{sd} & \textbf{coverage} & \textbf{power} & \textbf{mean} & \textbf{sd} & \textbf{coverage} & \textbf{power} \\
    \hline
    80\%  & 0.3   & 0.299  & 0.005  & 0.980  & 1     & 0.217  & 0.101  & 0.790  & 0.685 & 0.298  & 0.005  & 0.970  & 1     & 0.209  & 0.086  & 0.765  & 0.665 \\
          & 0.1   & 0.298  & 0.015  & 0.975  & 1     & 0.081  & 0.141  & 0.695  & 0.065 & 0.299  & 0.015  & 0.985  & 1     & 0.071  & 0.146  & 0.690  & 0.045 \\
    \hline
    50\%  & 0.3   & 0.300  & 0.004  & 0.975  & 1     & 0.245  & 0.118  & 0.920  & 0.580  & 0.299  & 0.004  & 0.980  & 1     & 0.265  & 0.113  & 0.935  & 0.595 \\
          & 0.1   & 0.302  & 0.013  & 0.960  & 1     & 0.169  & 0.277  & 0.925  & 0.115 & 0.302  & 0.013  & 0.955  & 1     & 0.122  & 0.268  & 0.900  & 0.075 \\
    \hline
    20\%  & 0.3   & 0.299  & 0.004  & 0.970  & 1     & 0.260  & 0.203  & 0.915  & 0.255 & 0.299  & 0.004  & 0.970  & 1     & 0.276  & 0.185  & 0.955  & 0.285 \\
          & 0.1   & 0.303  & 0.012  & 0.955  & 1     & 0.193  & 0.439  & 0.945  & 0.050  & 0.302  & 0.012  & 0.950  & 1     & 0.181  & 0.469  & 0.945  & 0.070 \\
    \hline
    \end{tabular}%
  \label{table1}%
\end{sidewaystable}

\begin{sidewaystable}[h!]
\scriptsize
  \centering
  \caption{Causal effects estimated from simulated data with continuous $Y_{1}$ and $Y_{2}$ using two different methods (our Bayesian method and classic IVW method) when $\beta_{1}=\beta_{2}=0$. Mean, standard deviation (sd), coverage under the null (coverage) and power are displayed for the estimated causal effects of the exposures $X_{1}$ and $X_{2}$ on the outcomes $Y_{1}$ and $Y_{2}$. There are 6 configurations containing three percentages of missing $X$ (80\%, 50\%, 20\%) and two levels of IV strengths ($\balpha = \mathbf{0.3}$ and $\mathbf{0.1}$). The effects of $U$ on all the exposures and outcomes are set to 1.}
    \begin{tabular}{|c|c|ccc|ccc|ccc|ccc|}
    \hline
    \multicolumn{1}{|c|}{\multirow{3}[3]{*}{\textbf{Missing rate}}} & \multirow{3}[3]{*}{\textit{$\balpha$}} & \multicolumn{6}{c|}{\textit{$\widehat{\beta_{1}}$}}           & \multicolumn{6}{c|}{\textit{$\widehat{\beta_{2}}$}} \\
\cline{3-14}          &       & \multicolumn{3}{c|}{\textbf{Bayesian}} & \multicolumn{3}{c|}{\textbf{Classic}} & \multicolumn{3}{c|}{\textbf{Bayesian}} & \multicolumn{3}{c|}{\textbf{Classic}} \\
\cline{3-14}          &       & \textbf{mean} & \textbf{sd} & \textbf{coverage} & \textbf{mean} & \textbf{sd} & \textbf{coverage} & \textbf{mean} & \textbf{sd} & \textbf{coverage} & \textbf{mean} & \textbf{sd} & \textbf{coverage} \\
    \hline
    80\%  & 0.3   & -0.001  & 0.005  & 0.960  & 0.007  & 0.061  & 0.955  & -0.001  & 0.005  & 0.955  & -0.005  & 0.062  & 0.960  \\
          & 0.1   & 0.004  & 0.016  & 0.960  & -0.010  & 0.112  & 0.965  & 0.004  & 0.015  & 0.960  & -0.001  & 0.130  & 0.960  \\
    \hline
    50\%  & 0.3   & 0.000  & 0.005  & 0.975  & -0.014  & 0.087  & 0.935  & 0.000  & 0.005  & 0.955  & -0.002  & 0.090  & 0.955  \\
          & 0.1   & 0.004  & 0.013  & 0.970  & 0.005  & 0.188  & 0.960  & 0.004  & 0.013  & 0.955  & -0.011  & 0.202  & 0.950  \\
    \hline
    20\%  & 0.3   & 0.000  & 0.004  & 0.950  & 0.010  & 0.148  & 0.930  & 0.000  & 0.004  & 0.965  & -0.003  & 0.152  & 0.935  \\
          & 0.1   & 0.003  & 0.012  & 0.965  & 0.012  & 0.394  & 0.920  & 0.003  & 0.012  & 0.965  & 0.020  & 0.361  & 0.945  \\
    \hline
    \end{tabular}%
  \label{table2}%
\end{sidewaystable}

\end{appendices}
}



\end{document}